\documentclass[conference]{IEEEtran}
\usepackage{geometry}
\usepackage{cite}
\usepackage{amsmath,amssymb,amsfonts}
\usepackage{graphicx}
\usepackage{textcomp}
\usepackage[dvipsnames]{xcolor}
\usepackage{soul}
\usepackage{url}
\usepackage{fancyhdr, lipsum}
\usepackage{setspace}

\usepackage{enumitem}
\usepackage[symbol]{footmisc}
\def\BibTeX{{\rm B\kern-.05em{\sc i\kern-.025em b}\kern-.08em
		T\kern-.1667em\lower.7ex\hbox{E}\kern-.125emX}}

\usepackage[utf8]{inputenc}
\usepackage[english]{babel}
\usepackage{amsthm}
\newtheorem{theorem}{Theorem}

\usepackage{verbatim}
\usepackage{tikz}
\usepackage{pgfplots}
\usepackage[normalem]{ulem}
\usepackage{algcompatible}
\usepackage{algpseudocode}
\usepackage{xcolor}
\usepackage{threeparttable}
\usepackage{pifont}
\usepackage{etaremune}
\usepackage{caption}
\usepackage{subcaption}
\usepackage[export]{adjustbox}
\usepackage{listings}
\usepackage{courier}

\usepackage{algorithm} 
\usepackage{algorithmicx}
\usepackage{fixltx2e}
\usepackage{comment}
\usepackage{multirow, booktabs}
\usepackage{siunitx}
\usepackage{dblfloatfix}
\usepackage{xargs}  
\usepackage{tikz}
\usetikzlibrary{matrix}
\usetikzlibrary{calc,fit}

\setlist{leftmargin=3.0mm}

\usepackage{tikz}
\usepackage{calc}

\IEEEoverridecommandlockouts
\begin{document}

\title{HiKonv: High Throughput Quantized Convolution With Novel Bit-wise Management and Computation\vspace{5mm}}

\author{
	\IEEEauthorblockN{
		Xinheng Liu$^{1*}$,
		Yao Chen$^{2*}$\thanks{*Both authors contribute equally to this paper.},
		Prakhar Ganesh$^{2}$,
		Junhao Pan$^{1}$,
		Jinjun Xiong$^{1,3}$,
		Deming Chen$^{1,2}$
	}
	\IEEEauthorblockA{
		$^{1}$University of Illinois at Urbana-Champaign, USA,
		$^{2}$Advanced Digital Sciences Center, Singapore, 
		$^{3}$University at Buffalo
	}
	\IEEEauthorblockA{
		\{xliu79, jpan22, jinjunx, dchen\}@illinois.edu, \{yao.chen, prakhar.g\}@adsc-create.edu.sg
	}
	\vspace{-10pt}
}

% use for special paper notices
%\IEEEspecialpapernotice{(Invited Paper)}

% make the title area
\maketitle

\makeatletter
\def\ps@IEEEtitlepagestyle{%
  \def\@oddfoot{\mycopyrightnotice}%
  \def\@evenfoot{}%
}
\makeatother
\def\mycopyrightnotice{%
  \begin{minipage}{\textwidth}
    \footnotesize
    ~ \hfill\\~\\
  \end{minipage}
  \gdef\mycopyrightnotice{}% just in case
}

{\small\bf Abstract---
Quantization for Convolutional Neural Network (CNN) has shown significant progress with the intention of reducing the cost of computation and storage with low-bitwidth data inputs. 
There are, however, no systematic studies on how an existing full-bitwidth processing unit, such as CPUs and DSPs, can be better utilized to carry out significantly higher computation throughput for convolution under various quantized bitwidths. 
In this study, we propose HiKonv, a unified solution that maximizes the compute throughput of a given underlying processing unit to process low-bitwidth quantized data inputs through novel bit-wise parallel computation.
We establish theoretical performance bounds using a full-bitwidth multiplier for highly parallelized low-bitwidth convolution, and demonstrate new breakthroughs for high-performance computing in this critical domain. For example, a single 32-bit processing unit can deliver 128 binarized convolution operations (multiplications and additions) under one CPU instruction, and a single 27$\times$18 DSP core can deliver eight convolution operations with 4-bit inputs in one cycle.
We demonstrate the effectiveness of HiKonv on CPU and FPGA for both convolutional layers or a complete DNN model.
For a convolutional layer quantized to 4-bit, HiKonv achieves a 3.17$\times$ latency improvement over the baseline implementation using C++ on CPU. Compared to the DAC-SDC 2020 champion model for FPGA, HiKonv achieves a 2.37$\times$ throughput improvement and 2.61$\times$ DSP efficiency improvement, respectively.}

\section{Introduction}\label{sec:intro}

Quantization is a frequently used technique in hardware implementation of Deep Neural Network (DNN) models
in order to reduce both the memory consumption and execution time~\cite{gholami2021survey, ul2q, vecq, skynet, DNNBuilder, cong2019dac}.
It is typically done by approximating high-precision floating point numbers to low-bitwidth integers or fixed-point numbers.
This is particularly important for modern DNN models as many of them employ convolutional layers, which contain intensive multiplication and accumulation (MAC)
operations~\cite{chen2019clouddnn, DNNBuilder, cong2019dac,tdla}.
Therefore, many novel quantization methods have been proposed in the literature to reduce the precision of weights, activations, and even gradients for DNNs while retaining their high accuracy~\cite{ul2q, vecq, cong2019dac, zhou2016dorefa}.

The current hardware implementation of quantized DNNs is, however, not ideal as there is no general support for quantized MACs without changing the underlying hardware~\cite{sharma2018bit,ryu2019bitblade,shin2018dnpu,lee2018unpu,sharify2018loom,pirdsp}.
Most hardware units have a high-bitwidth (such as 32 or 64 bits) MAC for either floating point numbers or integers~\cite{mliot}. When they are used
for quantized MACs, most of the bitwidths are left underutilized, wasting precious computing resources.
Even with the 8-bit multi-data processing of the Advanced Vector Extensions (AVX) support in X86\_64 architecture,
processing a single 4-bit multiplication would still occupy the 8-bit data width with
the remaining 4 bits simply wasted~\cite{AVXforquant}.
The waste becomes even more severe when either processing lower bitwidth (such as binary) data or utilizing a hardware unit with higher built-in bitwidths.

Reconfigurable hardware such as FPGA may alleviate some of the waste because of its bit-level flexibility for configuration, but it
exhibits similar drawbacks, especially for FPGAs with high-precision Digital Signal Processing (DSP) units~\cite{mliot, tdla, chen2016platform}.
Without a careful bit-wise management of inputs and outputs, deploying quantized DNNs onto FPGAs with the given DSPs
would still waste much of their computation capacity.

In this paper, we propose a novel solution, HiKonv, that can significantly improve the existing arithmetic units' utilization
efficiency when conducting quantized convolution, thus improving throughput for convolution and reducing end-to-end DNN computation latency.
Our solution is based on a careful management of bitwidths used for quantized MACs and novel mapping of multiple parallelized MACs onto an existing
arithmetic unit, such that the arithmetic unit's computation capacity is fully utilized. We further show theoretically that such a management and mapping strategy is universal in the sense that it can be applied to arbitrarily quantized bitwidths and high-bitwidth arithmetic units. For example,  a  single 32-bit processing  unit  can  deliver  128  binarized  convolution operations using one instruction for CPU, and a single  27$\times$18 DSP core can deliver eight convolution operations with 4-bit inputs in one cycle.
Based on such a theoretical analysis, we show that there are different optimal design points in choosing the quantization bitwidth for a
given arithmetic processing unit. Our experimental results further validate our analysis and HiKonv's general applicability. For example, our
CPU-based implementation of HiKonv achieves up to $3.17 \times$ latency improvement for quantized convolution over existing methods on the same CPU.
We also apply HiKonv to an end-to-end quantized DNN model, UltraNet~\cite{UltraNet2020}, in an FPGA setting, and the measured
on-board result outperforms the state-of-the-art in terms of throughput and DSP efficiency by $2.37 \times$ and $2.61 \times$, respectively.

Because of its generality, we believe HiKonv opens up a new venue for further improving the hardware efficiency of DNN based inferences. It
not only improves the throughput and latency for existing quantized DNN models on existing hardware, but also offers new opportunities
for designing new hardware-friendly quantized DNN models or co-designing both the hardware and quantized DNN models.

% \red{JX: we can include Figure 1 in Section II.A when discussing Input Slicing}

\section{Preliminary}\label{sec:prelim}

Before we present our proposed HiKonv solution, we first review 1) input slicing and data packing for concurrency improvement and 2) 1-D convolution.

\subsection{Input-Slicing for Concurrency Improvement}

\begin{figure}[h]
    \centering
    \includegraphics[width=0.48\textwidth]{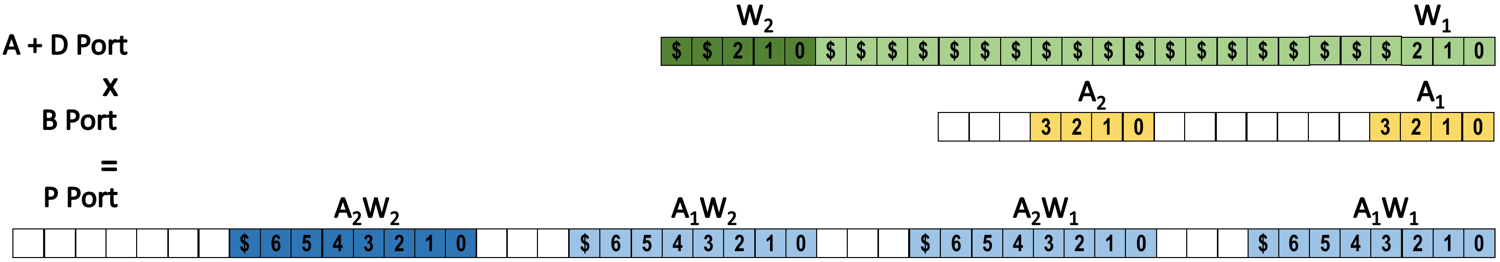}
    \caption{INT4 optimization on DSP48E2~\cite{xilinxint4}.}
    \label{fig:xilinxint4}
\end{figure}

Input slicing and data packing are
% a novel method that first created by FPGA vendors to support low-bitwidth data processing.
generally used by the current hardware units to process low-bitwidth inputs~\cite{xilinxint4, xilinxint8, AVXforquant}.
The input bitwidth is split into different pieces, each of which is called a slice to hold a low-bitwidth data.
It uses the available bitwidth space to hold a number of data slices to improve the parallelism while still preserving the correct output.
An example of INT4 optimization for Xilinx DSP48E2 unit is shown in Figure~\ref{fig:xilinxint4}, each of the input contains two slices.
\vspace{-5pt}
It takes advantage of the multiple input ports and the internal addition operation of the DSP to enable four multiplications of data slices simultaneously.
% More specifically, each of the port to the multiplier contains two 4-bit data.
Since an INT4$\times$UINT4 multiplication generates a result that needs at least an 8-bit space,
guard bits are added during the packing of the low-bitwidth data to guarantee the correctness of the result. 
Here, we define the term \textit{guard bit} as the filling 1s or 0s between the packed data in the multiplicand for the purpose of preventing computation overflow. 
% (This part is about the Figure 1, how could we get the guard bits as 3 here? in the following sentence.)}
% When four channels of MAC are packed together, enough guard bits need to be placed between the two inputs. 
% In this INT4 optimization, according to the DSP48E2's input sizes and internal logic, 
% \textcolor{blue}{Particularly, the length of guard bits is set to three according to the DSP48E2's input size.(We need a clearer definition for guard bits, and also 3 seems not correct here.)}
% \textcolor{red}{Xinheng: Just delete this sentence, not important information}
The multiplication with the sliced and packed inputs is represented as:
\begin{equation}
\label{equ:xilinx4bit}
\small
\begin{split}
    &(A_2 \cdot 2^{11} + A_1) \cdot (W_2 \cdot 2^{22} + W_1) \\
    &= A_2W_2\cdot 2^{33} + A_1W_2\cdot2^{22} + A_2W_1\cdot2^{11} + A_1W_1
\end{split}    
\end{equation}
The output of Equation~\ref{equ:xilinx4bit} is the concatenation of four multiplication results with zeros between them due to the guard bits. This process accomplishes four multiplications within one operation cycle.
% Nevertheless, \cite{AVXforquant} takes advantage of the 8-bit multi-data processing of the AVX support in X86\_64 architecture and packs 4-bit data into each of the 8-bit spaces.

\subsection{1-D Convolution}
The conventional 1-D discrete convolution between an $N$-element sequence $f$ and a $K$-element kernel $g$ (denoted as $F_{N,K}(f,g)$) can be represented as Equation \ref{eq:pad_f}. Here, we define the infinite length sequence $h$ as the zero extension of $f$ with the index range of $(-\infty,\infty)$. Meanwhile, $y$ is the output with $N\text{+}K\text{-}1$ non-zero elements.

\begin{small}
\begin{gather}
\label{eq:pad_f0}
    h[n]=\begin{cases} 
        f[n] &,  0\leq n < N\\
        0 &, n <0 \text{ or } n\geq N
    \end{cases}\\
    y[m]= (h\ast g)[m]=\sum_{k=0}^{K-1}h[m-k]g[k] \label{eq:pad_f}
\end{gather}
\end{small}
Alternatively, $y$ can be represented as an $(N\text{+}K\text{-}1)$-element sequence with Equation~\ref{eq:ym}. Each of the $y[m]$ involves a sequence of multiplication and addition operations. 
\begin{equation}
\small
\label{eq:ym}
y[m]=\sum_{k+n=m} h[n]g[k]
\end{equation}

% Meanwhile, the 1-D convolution has the distributive property and the shift property regarding the zero-padded sequence $h$ from $f$: 
% \begin{gather}
%     h_1 * g+h_2*g=(h_1+h_2)*g\\
%     (h*g)[n-n_0]=(h'*g)[n], h'[n]=h[n-n_0] 
% \end{gather}

% The convolution operation $*$ between $h$ and $g$ in Equation \ref{eq:pad_f} has the distribution and shifting property:
% \begin{gather}
%     h_1*g+h_2*g=(h_1+h_2)*g\\
%     (h*g)[n-n_0]=(h_{n_0}*g)[n], h_{n_0}[n]=h[n-n_0]
% \end{gather}

% 1-D convolution is the basic operation that forms the 2-D and DNN convolutions.

\section{Multiplier for Convolution}\label{sec:formula}

% \textcolor{blue}{(One change I have to do here is about the sign bit operations. We should maintain the generalized presentation in this section because the technique is a general technique for existing hardware and new hardware. So we do not need to talk about the hardware design overhead here, move them to the final section for hardware construction.)}

\begin{figure*}
    \centering
    % \vspace{-56pt}
    \includegraphics[width=0.85\textwidth]{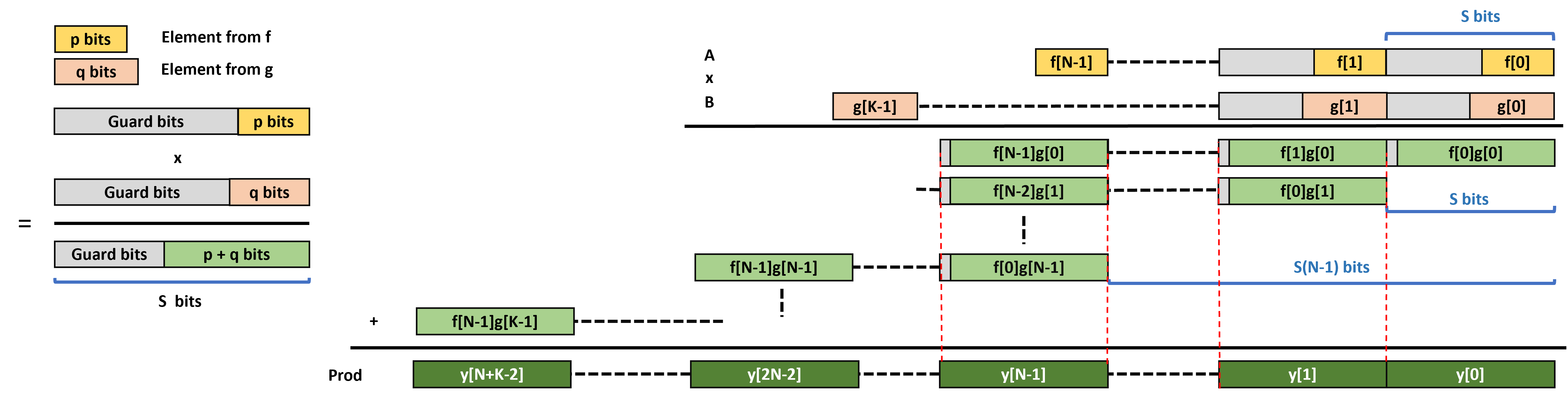}
    \caption{Ideal process of $Prod = A\times B$.}
    \vspace{-12pt}
    \label{fig:amulb}
\end{figure*}

Inspired by input slicing and data packing for novel bit management and high processing concurrency, we generalize the solution for using a given hardware unit to process the maximum amount of low-bitwidth convolution operations concurrently with theoretical guarantees.

We first define the variables related to our exploration.
As shown in Figure~\ref{fig:amulb}, we assume a given high-precision hardware unit that can multiply $Bit_A$-bit integer input $A$ with $Bit_B$-bit integer input $B$ and generate the product $Prod$. The bitwidths of $A$ and $B$ define the computation capability of the hardware unit, or more specifically, the multiplier, and thus determine the design setting of HiKonv specific to this unit. Convolution input $f$ and kernel $g$ are the two sequences of low-bitwidth integer values quantized to $p$ and $q$ bits, respectively.
% Elements in $f$ and $g$ are quantized to $p$ and $q$ bits, respectively.
% Here, for this given hardware unit with $A$ and $B$ as inputs, we want to know what is the maximum number of multiplications we could achieve between the items from $f$ and $g$.

To determine how to load $A$ and $B$ with multiple convolution operands from $f$ and $g$ and perform the convolution between these operands, we define an additional variable $S$ to be the size of a slice of inputs for both $A$ and $B$ as demonstrated on the left in Figure~\ref{fig:amulb}.
% mentioned in Section \ref{sec:prelim}. 
The lower bits of each slice contain one operand from $f$ or $g$.
% \textcolor{blue}{The size of the slice $S$ determines the number of bits to shift left between the intermediate results' slices from different stages and different adjacent products of elements from $f$ and $g$.}
To simplify the problem, we assume the $N$ and $K$ are the maximum numbers of operands from $f$ and $g$ that can fit into $A$ and $B$, respectively.
Hence, the polynomial representations of $A$ and $B$ are:

\begin{small}
\begin{equation}\label{eq:inpack}
% \begin{split}
        A=\sum_{n=0}^{N-1} f[n]\cdot 2^{ S\cdot n }, ~~
    B=\sum_{k=0}^{K-1} g[k]\cdot 2^{ S\cdot k }
% \end{split}
\end{equation}
\end{small}

Although the intermediate results of the multiplication are invisible to us, we assume the processing unit takes the most ideal way for the multiplication of two inputs, as shown in Figure~\ref{fig:amulb}. 
The entire multiplication is treated as the multiplication of slices in $A$ with the corresponding slices in $B$ followed by shifting the product left by $S$ bits and accumulating the shifted results to the previous result.
There are always $N\times K$ products between elements from $f$ and $g$ that are computed and accumulated to form the output $Prod$.

\subsection{From Multiplication To Convolution}
\label{subsect:mul_conv}

To use the product $Prod =A \times B$, we need to segment the output into an effective format for convolution during the process.
In order to segment the intermediate results, we extend the guard bits $G_b$ in~\cite{xilinxint4}.
% to avoid any overlapping of the effective products between the adjacent intermediate segments in the same row as well as the accumulated segments vertically.
The guard bits are not only to prevent overlaps between the effective product of two adjacent intermediate partial products but also to segment out the partial accumulations of vertically stacked segments.
Its length varies according to the maximum number of multiplication terms $f[n]g[k]$ that are summed together. 
According to our settings for $A$ and $B$, a maximum of $min(K,N)$ terms are summed together for each output segments. 
% Accordingly, the guard bit $G_b=\lceil{log_2 min(K,N)}\rceil$ in the single $A\times B$ multiplication. 
% With the variables defined above, we could know that 
Therefore, to ensure the correctness of the final result, each of the slicing should contain both the guard bits and the bits of the $p$-bit and $q$-bit inputs for the production, respectively.
% the slicing size has to be big enough to avoid overflow for the multiplication of $f[n]g[k]$ and also include the guard bits to avoid the overflow caused by the accumulations happen vertically. 

\begin{theorem}\label{lemma:seg}
Assuming guard bits, $G_b$, are properly decided according to the specific multiplier setting, with given $A$ and $B$ constructed from $N$-element sequence $f$ and $K$-element sequence $g$, where $f$ and $g$ are quantized respectively to $p$ and $q$ bits, we can obtain $N+K-1$ segments from the product $Prod =A\times B$ which are all short partial convolutions in the form of 1-D convolution.
\end{theorem}
\begin{proof} Considering the guard bits, we can obtain:
\begin{small}
\begin{gather}\label{equ:slice1}
    S = \begin{cases}
     q+G_b, &p=1, q\geq 1\\
     p+G_b, &q=1, p\geq 1\\
     p+q+G_b, &otherwise
    \end{cases}
\end{gather}
\end{small}
\vspace{-0.5em}
\begin{small}
\begin{equation}\label{equ:slice2}
p+(N-1)S\leq Bit_A
\end{equation}
\end{small}
\vspace{-1.5em}
\begin{small}
\begin{equation}\label{equ:slice3}
q+(K-1)S\leq Bit_B
\end{equation}
\end{small}
Thereby, the intermediate stages are shifted left by $S$ bits for every stage, and the effective vertical accumulation of the partial products in the segments from all the stages stacked together would not exceed the length of $S$ bits, as shown in Figure~\ref{fig:amulb}. Then, the multiplication is represented as:
\begin{equation}
\small
\begin{split}
    Prod &=A \times B
    =(\sum_{n=0}^{N-1} f[n]\cdot2^{ S\cdot n })\cdot(\sum_{k=0}^{K-1} g[k]\cdot 2^{ S\cdot k })\\
    &=\sum_{m=0}^{N+K-2} ( \sum_{n+k=m}(f[n]\cdot2^{ S\cdot n }\cdot g[k]\cdot 2^{ S\cdot k }) )\\
    &=\sum_{m=0}^{N+K-2}(\sum_{n+k=m}(f[n] \cdot g[k]) \cdot 2^{S\cdot m})
\end{split}
\end{equation}
% Overall, the number of segmented out data is $N+K-1$ and there will be $N \times K$ effective multiplication happens between the elements from $f$ and $g$ sequences with regardless of obtaining each of the multiplication results for a pair of elements out when a single result of $A \times B$ produced. 
Different from general multiplications, convolution consists of a sequence of multiplications and accumulations.
Referring to the form of 1-D convolution in Equation~\ref{eq:ym}, the result of $Prod$ can be represented as:
\begin{equation}
\small
    Prod =\sum_{m=0}^{N+K-2} y[m]\cdot 2^{ S\cdot m}     
\end{equation}
where the intermediate accumulations form a 1-D convolution of two sequences in each of the output segments, and
% with the number of elements ranging from 1 to $N$ and 1 to $K-1$, respectively. \textcolor{red}{What is the number/element defined here?} 
the total number of convolution segments is $N+K-1$.
\end{proof}

% \paragraph{Input slicing and output segmentation}
% \subsection{Dealing With Sign Bit}\label{subsect:sign_bit}
Per the above, we can use a high-bitwidth multiplier to process two integers $A$ and $B$ to form $N+K-1$ convolutions of short sequences.
% However, the representation of signed and unsigned values affect the final value of $A$ and $B$. 
However, due to the two's complement representation of signed values, directly packing negative values into $A$ or $B$ leads to wrong results for the intermediate products. 
To guarantee the correctness of the products as the intermediate results, we must consider the sign bit during the packing of the elements from $f$ and $g$ into $A$ and $B$ as well as segmenting the result $Prod$.

% \red{JX: Move the unsigned equation to the previous section?}
% \blue{XL: I think the two section can be merged}
% \textcolor{blue}{(Update the following part according to Xinheng's presentation slides. Revise the equations and add in the new figure.)}

If $f$ and $g$ are all unsigned integers, we can construct $A$ and $B$ as integers with bit-wise assignments and the zero extension without additional operations:
% \textcolor{red}{without hardware overhead}:
\begin{equation}
\small
\begin{split}
    A[S(n\text{+}1)\text{-}1\text{:}Sn]&= f[n]\\
    B[S(k\text{+}1)\text{-}1\text{:}Sk]&= g[k]
\end{split}
\end{equation}
Meanwhile, each $y[m]$ can be segmented out from $Prod$ with:
\begin{equation}
\small    
    y[m]= Prod[S(m\text{+}1)\text{-}1\text{:}Sm]
\end{equation}
However, if $f$ and $g$ contain signed integers, we need additional bit management.

\begin{figure}[h]
    \centering
    \includegraphics[width=0.48\textwidth]{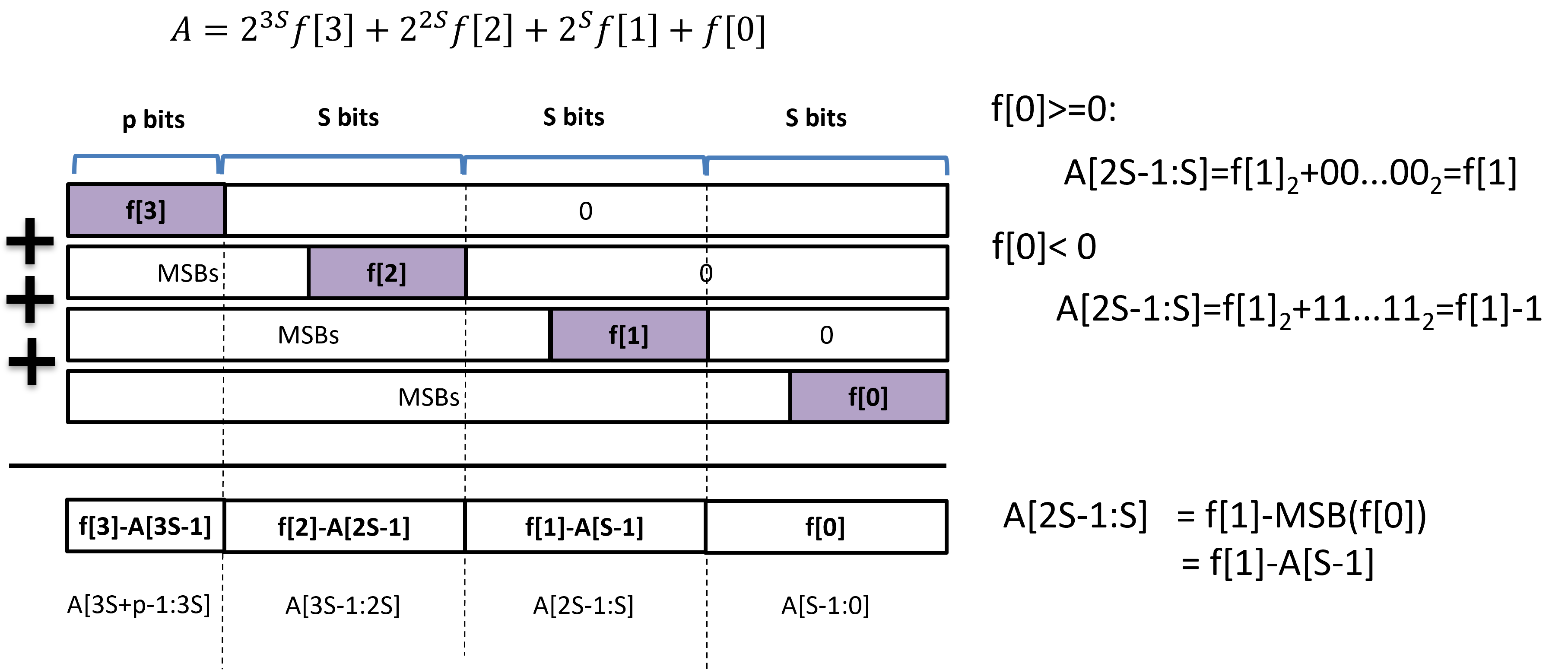}
    \vspace{-6pt}
    \caption{Input packing for signed integer $f$ sequence}
    \label{fig:signpack}
\end{figure}
\vspace{-6pt}

Figure \ref{fig:signpack} shows the packing of four elements of $f$ into multiplicand $A$.
Taking the second $S$-bit segment as an instance, in 2’s complement expression, if $f[0]$ is positive, the MSB is 0, and the sign extension part are all zeros.
On the other hand, if $f[0]$ is negative, the sign extension part are all $1$s in binary expression and represents -1 in 2’s complementary representation.
In such condition, we decrement 1 from $f[1]$ to form the second S-bit 
and perform the packing process with concatenation and 1-bit incrementer instead of using a larger bitwidth adder.
% in hardware.}
The packing process works recursively for all the slices
while slicing of the output works in a reversed manner.
Equation \ref{eq:signpack} shows the packing and slicing formula for signed integer $f$ and $g$. With this bit management technique, we obtain $N + K -1$ partial convolutions from $N \times K$ segments of intermediate results for a single multiplier to process signed input data.

\vspace{-5pt}
\begin{equation}\label{eq:signpack}
\small
\begin{split}
      A[S(n\text{+}1)\text{-}1\text{:}Sn]= \begin{cases}
    f[0] &, n=0\\
    f[n]\text{-}A[Sn\text{-}1]&, n>0
    \end{cases}\\
    B[S(k\text{+}1)\text{-}1\text{:}Sk]= \begin{cases}
    g[0] &, k=0\\
    g[k]\text{-}B[Sk\text{-}1]&, k>0
    \end{cases}\\
    y[m]= \begin{cases}
    Prod[S\text{-}1\text{:}0] &, m=0\\
    Prod[S(m\text{+}1)\text{-}1\text{:}Sm]\text{+}Prod[Sm\text{-}1]&, m>0
    \end{cases}
\end{split}
\end{equation}

\subsection{Convolution Extension}
% Due to the variation of the value of $N$, $K$, we may not be able to directly use the results for our required convolution.
% We now look into the details of adjusting the outputs to form a normal convolution.

% \paragraph{Matching to convolution}
% As we can see from the Figure~\ref{fig:amulb}, 

% As shown in Figure~\ref{fig:amulb}, the invisible process to generate $P$ has an internal shift-and-accumulation pattern that forms a parallelogram. The segments in the middle of $P$ contains more number of multiplication results, and the segments on the sides have gradually reduced number of these results.
Now we have presented an efficient algorithm to use the multiplication unit on a hardware platform to perform the $F_{N,K}$ 1-D convolution. 
However, the size of $N$ are limited by the bitwidth of the hardware multiplier whereas most real-world applications have much larger input sizes.
Moreover, the $F_{N,K}$ 1-D convolution is often used as a unit building block for other larger-scale convolution operations. 
Thus, we design a new algorithm to use the $F_{N,K}$ 1-D convolution to complete arbitrarily large size 1-D convolutions and any arbitrary convolutions.
% \textcolor{blue}{\st{with the aid of hardware accumulators}}.
As shown in Figure~\ref{fig:amulb}, the order of the elements for these intermediate production is controlled by the order of elements packed into the slices in $A$ and $B$;
it allows us to devise different accumulation methods to provide flexibility to construct different convolutions beyond 1-D convolution. 

\begin{figure}[h]
    \centering
    \includegraphics[width=0.4\textwidth]{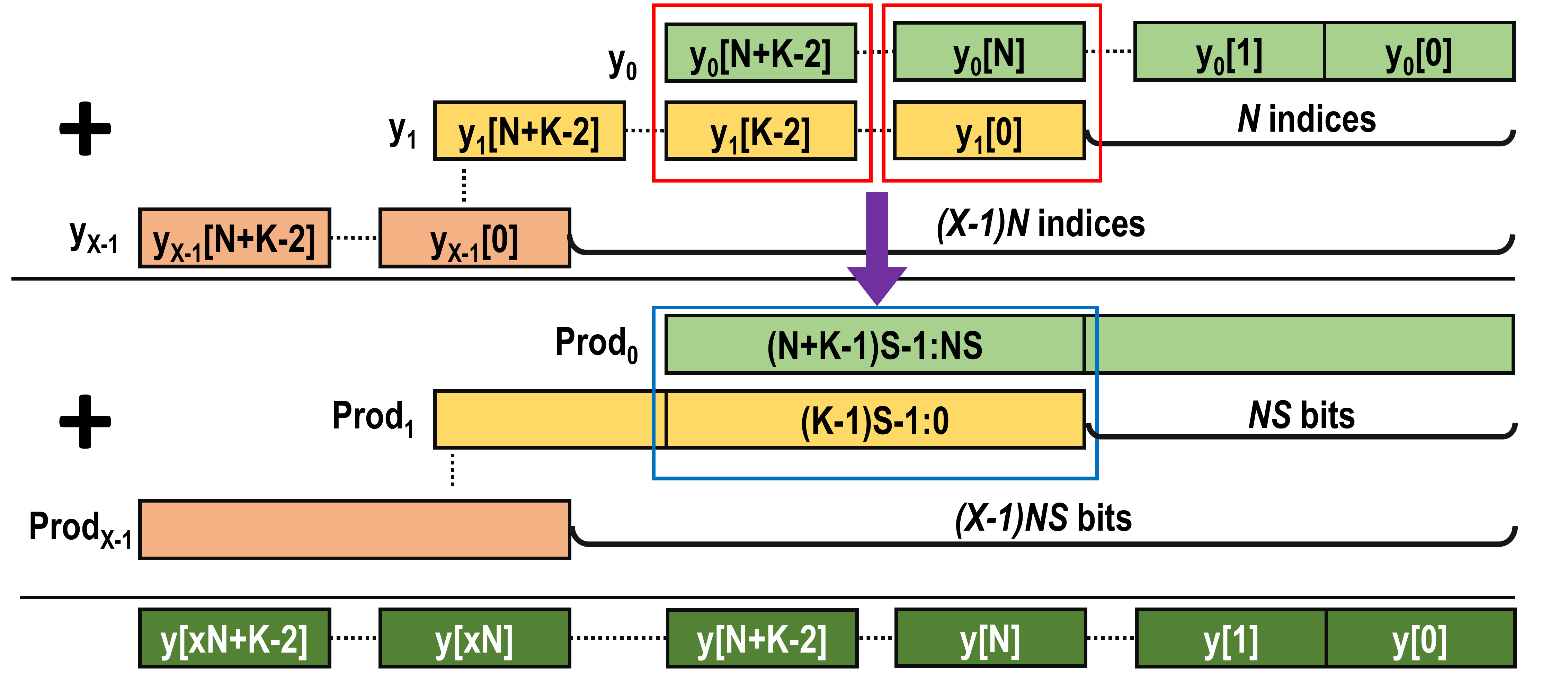}
    \caption{Computation of $F_{XN,K}$ 1-D convolution}
    \label{fig:horistack}
\end{figure}

\paragraph{1-D Convolution Extension} 
% \st{Section} \ref{subsect:mul_conv} \st{presents the approach to compute a 1-D $F_{N,K}$ convolution of integer sequences using one multiplication unit.}
Regarding the $F_{N,K}$ as a basic operation, we extend it to $F_{X\cdot N,K}$ convolution of a longer sequence by summing up the elements in output sequences of different $F_{N,K}$ convolutions. 
\begin{theorem}\label{lemma:h_stack}
The output sequence $y=F_{X\cdot N,K}$ of a 1-D convolution between an $(X\cdot N)$-element sequence $f$ and a $K$-element filter $g$ can be represented as the sum of index-shifted output sequences $y_x=F_{N,K}(f_x,g)$, as shown in Equation \ref{eq:1dnx}. Here, $f_x=f[xN\text{:}(x\text{+}1)N\text{-}1]$ ($x \in [0, X-1]$).
\end{theorem}
\begin{proof} 
Following Equation \ref{eq:pad_f0}, we extend $f$ and $f_x$ sequences into zero extension sequences $h$ and $h_x$. Then $h$ is represented as the sum of index-shifted $h_x$ sequence:
\begin{equation}
\small
    h[n]= \sum_{x=0}^{X-1} h_x[n-xN]
\end{equation}
According to Equation \ref{eq:pad_f}, the convolution output $y$ is calculated with:
\begin{equation}
\small
\begin{split}
    y[n]&=\sum_{k=0}^{K-1}h[n-k]g[k]\\
    &=\sum_{k=0}^{K-1}(\sum_{x=0}^{X-1} h_x[n-xN-k])g[k] \\
    &=\sum_{x=0}^{X-1}( \sum_{k=0}^{K-1}h_x[n-xN-k]g[k])
\end{split}
\end{equation}
Given that $y_x[n]\text{=}\sum_{k\text{=}0}^{K\text{-}1}h_x[n\text{-}k]g[k]$, we can represent sequence $y$  as the sum of index-shifted $y_x$ sequences.
\begin{equation}
\label{eq:1dnx}
\small
    y[n]=\sum_{x=0}^{X-1} y_x[n-xN]
\end{equation}
\end{proof}
\vspace{-6pt}
Equation \ref{eq:1dnx} reveals that the extended $F_{X\cdot N,K}$ 1-D convolution is computed by a shift-accumulation pattern with $F_{N,K}$ base operation results. 
Figure \ref{fig:horistack} demonstrates how the elements in different $y_x$ sum up to the elements in $y$. Each computed $y_x$ sequence is shifted $xN$ indices and then summed up to form the element of $y$, which is marked by the red square.
Still, we use existing adder unit of the given platform to complete multiple additions by adding the bit slices from the product $Prod$ as mentioned in Section \ref{subsect:mul_conv} as marked by the blue square. 
In such a case, the guard bit of $G_b = \lceil{log_2K}\rceil$ is also adjusted with additional bits to prevent the partial results from overflow.

%   \begin{gather}
% \label{eq:nd_conv}
%     z[iN\text{:}i(N\text{+}1)\text{-}1]=\begin{cases} 
%         [y^{0}_0,y^{0}_1] &, i=0\\
%         [y^{N-1}_2,y^{N-1}_3] &, i=N\\
%         [y^i_0+y^{i-1}_{2},y^i_1+y^{i-1}_{3}] &, other wise 
%     \end{cases}
% \end{gather}
        
\paragraph{DNN Convolution}

% The convolution layer in DNN computes a feature-map array $I[C_i][H_i][W_i]$ and a kernel array $W[C_o][C_i][K][K]$ for output feature-map array $O[C_o][H_o][W_o]$ which can be represented as:
% \begin{equation}
% \small
%   O[c_o][h][w]\text{=}\sum_{c_i\text{=}0}^{C_i\text{-}1}\sum_{k_h\text{=}0}^{K\text{-}1}\sum_{k_w\text{=}0}^{K\text{-}1}I[c_i][h\text{+}k_h][w\text{+}k_w]W[c_o][c_i][k_h][k_w]  
% \end{equation}
% With the inherent convolution computation pattern, we may also   compute a DNN convolution layer with $F_{N,K}$ 1-D convolution as the base operation, as shown in Theorem \ref{thrm:convolution}. 
% The proof is not included due to the page limit.  
% \begin{theorem}
% \label{thrm:convolution}
% For a DNN convolution, the output feature-map can be computed by $F_{N,K}$ 1-D convolution with the following equation:
% \begin{equation}
% \small
%     O[c_o][h][w]=\sum_{c_i\text{=}0}^{C_i\text{-}1}\sum_{k_h\text{=}0}^{K\text{-}1}\sum_{x\text{=}0}^{\lceil{\frac{W_o}{N}}\rceil} y_{c_i,c_o,h,k_h,x}[w\text{-}xN\text{+}K\text{-}1]
% \end{equation}
% where 
% \begin{equation}
%     \small
%     \begin{cases}
%     f=I[c_i][h+k_h][xN\text{:}(x\text{+}1)N\text{-}1]\\
%     g=W[c_o][c_i][k_h][K\text{-}1\text{:}0]\\
%     y_{c_i,c_o,h,k_h,x}\text{=}F_{N,K}(f, g)
%     \end{cases}
% \end{equation}

% \end{theorem}
The convolution layer in DNN computes a feature-map array $I[C_i][H_i][W_i]$ and a kernel array $W[C_o][C_i][K][K]$ for output feature-map array $O[C_o][H_o][W_o]$ (assuming $H_i\text{=}H_o\text{+}K\text{-}1$ and $W_i\text{=}W_o\text{+}K\text{-}1$) which can be represented as:
\begin{equation}
\label{eq:conv_org}
\small
\begin{split}
   O[c_o]&[h][w]\text{=} \\
   &\sum_{c_i\text{=}0}^{C_i\text{-}1}\sum_{k_h\text{=}0}^{K\text{-}1}\sum_{k_w\text{=}0}^{K\text{-}1}I[c_i][h\text{+}k_h][w\text{+}k_w]W[c_o][c_i][k_h][k_w]  
\end{split}
\end{equation}
With the inherent convolution computation pattern, we may also  compute a DNN convolution layer with $F_{N,K}$ 1-D convolution as the base operation, as shown in Theorem \ref{thrm:convolution}. 

\begin{theorem}
\label{thrm:convolution}
For a DNN convolution, the output feature-map can be computed by $F_{X\cdot N,K}$ 1-D convolution with the following equation:
\begin{equation}
\small
\label{eq:dnn_hikonv}
    O[c_o][h][w]=\sum_{c_i\text{=}0}^{C_i\text{-}1}\sum_{k_h\text{=}0}^{K\text{-}1} y_{c_i,c_o,h,k_h}[w\text{+}K\text{-}1]
\end{equation}
Here, the term $y_{c_i,c_o,h,k_h}$ is a 1-D convolution result with $X=\lceil{\frac{W_i}{N}}\rceil-1$:
\begin{equation}
\small
\label{eq:conv_seg}
y_{c_i,c_o,h,k_h}=F_{X\cdot N,K}(f, g) 
\end{equation}
where $f$ and $g$ are defined as:
\begin{equation}
\small
\label{eq:I_PRIME}   
\begin{split}
f[w]&=\begin{cases} I[c_i][h\text{+}k_h][w], & 0\leq h < H_i, 0\leq i < W_i\\
0, & otherwise
\end{cases}\\
g&=W[c_o][c_i][k_h][K\text{-}1\text{:}0]
\end{split}
\end{equation}
\end{theorem}
\begin{proof}
For abbreviation, we denote sequence $y_{c_i,c_o,h,k_h}$ as $y'$. According to the definition of 1-D convolution, sequence $y'$ can be computed by following equation:

\begin{equation}
\small
\label{eq:o_partial2}
\begin{split}
    y'[n]&=\sum_{k=0}^{K-1}f[n\text{-}k]g[k]\\
    &=\sum_{k=0}^{K\text{-}1}I[c_i][h\text{+}k_h][n\text{-}k]W[c_o][c_i][k_h][K\text{-}1\text{-}k]\\
    &=\sum_{k=0}^{K\text{-}1}I[c_i][h\text{+}k_h][n\text{+}k\text{-}K\text{+}1]W[c_o][c_i][k_h][k]
\end{split}
\end{equation}
Then we have
\begin{equation}
\small
\label{eq:off}
        y_{c_i,c_o,h,k_h}[n\text{+}K\text{-}1]=\sum_{k=0}^{K\text{-}1}I[c_i][h\text{+}k_h][n\text{+}k]W[c_o][c_i][k_h][k]
\end{equation}
With Equation \ref{eq:off}, Equation \ref{eq:conv_org} could be represented as:

\begin{equation}
\small
\label{eq:1dfor3d}
\begin{split}
       O&[c_o][h][w] \\
       &\text{=}\sum_{c_i\text{=}0}^{C_i\text{-}1}\sum_{k_h\text{=}0}^{K\text{-}1}\sum_{k_w\text{=}0}^{K\text{-}1}I[c_i][h\text{+}k_h][w\text{+}k_w]W[c_o][c_i][k_h][k_w]\\ 
       &\text{=}\sum_{c_i\text{=}0}^{C_i\text{-}1}\sum_{k_h\text{=}0}^{K\text{-}1}(\sum_{k\text{=}0}^{K\text{-}1}I[c_i][h\text{+}k_h][w\text{+}k]W[c_o][c_i][k_h][k])\\
       &\text{=}\sum_{c_i\text{=}0}^{C_i\text{-}1}\sum_{k_h\text{=}0}^{K\text{-}1}y_{c_i,c_o,h,k_h}[n\text{+}K\text{-}1]
\end{split}    
\end{equation}
% Equation \ref{eq:dnn_hikonv} from 
\end{proof}
% \vspace{-10pt}
A convolution layer in DNN has multiple input and output channels, which require accumulation of channel-wise features to form the final output. 
By grouping the $F_{N,K}$ output sequences with different $c_i$ but same $c_o$,$h$,$k_h$ and $x$ indices, and accumulating the corresponding product $Prod$ with adders, we can perform the channel-wise accumulation of the feature maps. 
In this case, the required number of guard bits is $G_b=\lceil{log_2(M\cdot min(K,N))}\rceil$ for the accumulation of $M$ feature maps along input channel in a convolution.

% \begin{figure}
%     \centering
%     \includegraphics[width=0.3\textwidth]{figures/verticalStack2.png}
%     \caption{}
%     \vspace{-10pt}
%     \label{fig:vertistack}
% \end{figure}

% due to the accumulation happen for vertical segments.
% Overall, different stacking methods provide the flexibility to form different kinds of convolutions.
\subsection{Throughput Analysis}
Based on the above discussions,
the equivalent achievable throughput for convolution of inputs with $p$ and $q$ bits quantized data by a given processing unit is a function of both the supported bitwidth of $A$ and $B$ by the hardware and the given bitwidth of the elements in $f$ and $g$. It can be represented by the number of multiplication and accumulation operations ($\#$ ops) that one multiplier can perform for the low-bitwidth data in every cycle, which is $N \times K + (N-1) \times (K-1)$.

\begin{figure}[ht]
\begin{subfigure}{0.24\textwidth}
    \centering
    \includegraphics[width=\linewidth]{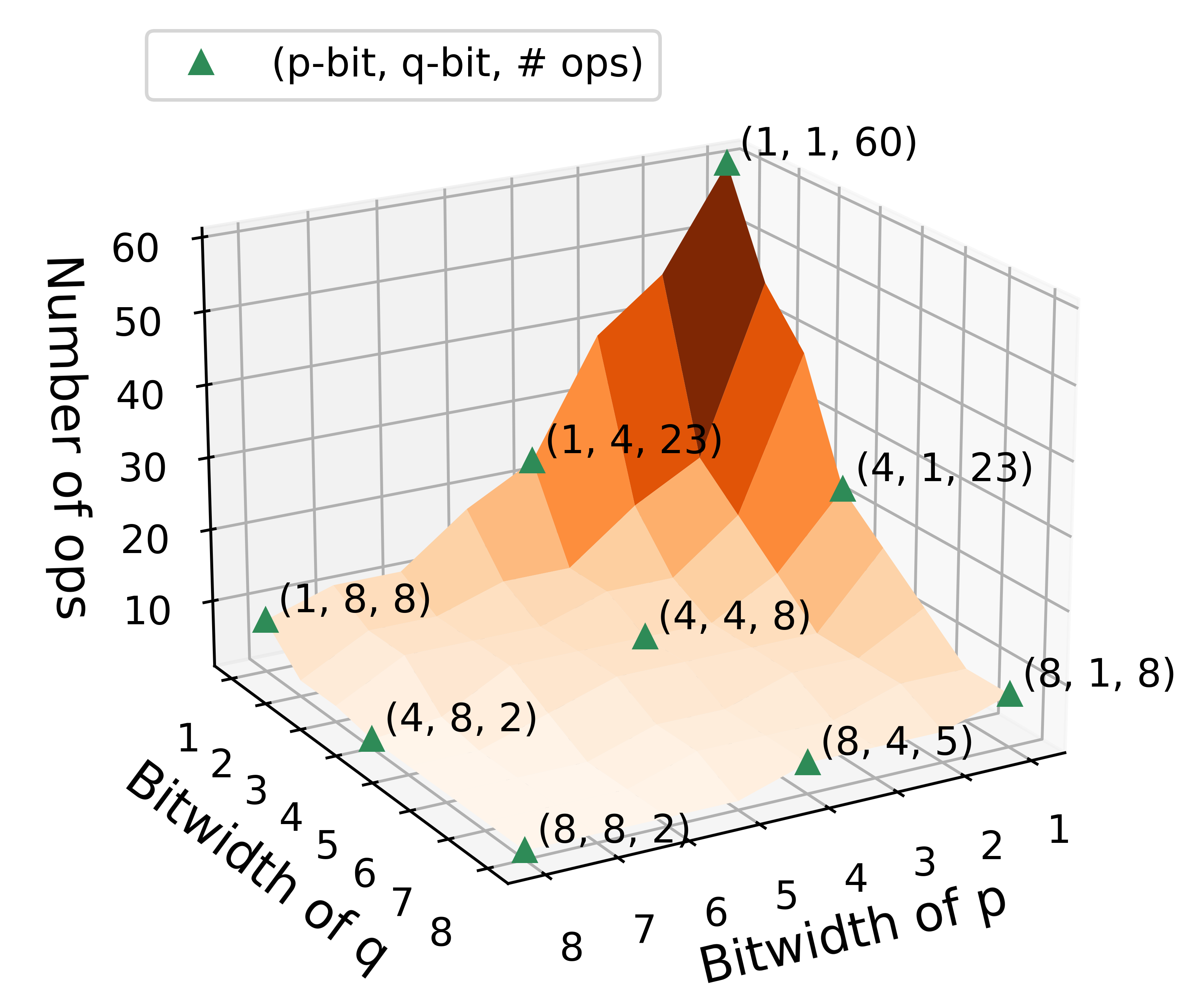}
    \caption{A = 27 bits, B = 18 bits}
    \label{fig:maxMul27}
\end{subfigure}%
\begin{subfigure}{0.24\textwidth}
    \centering
    \includegraphics[width=\linewidth]{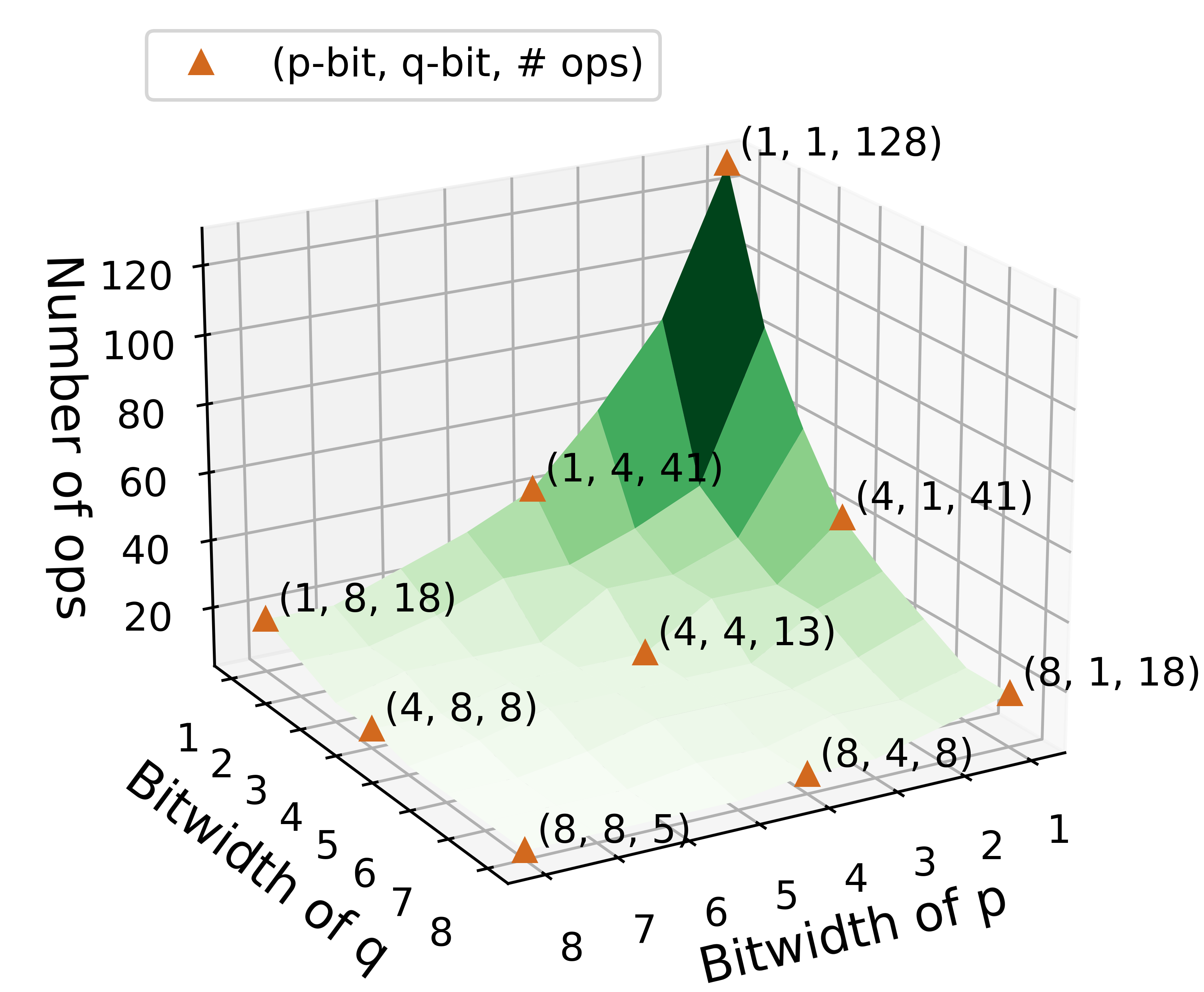}
    \caption{A = 32 bits, B = 32 bits}
    \label{fig:maxMul32}
\end{subfigure}
% \centering
% \subfloat[A = 27 bits, B = 18 bits]{%
%   \includegraphics[clip,width=0.8\columnwidth]{figures/maxmul27.png}%
%   \label{fig:maxMul27}
% }

% \subfloat[A = 32 bits, B = 32 bits]{%
%   \includegraphics[clip,width=0.8\columnwidth]{figures/maxmul32.png}%
%   \label{fig:maxMul32}
% }
\caption{Throughput of processing units with different bitwidth settings. The number of operations (z-axis) can be calculated by solving for $N$ and $K$ with $p$ and $q$ bits inputs, which are specific to the processing unit settings.}
% \textcolor{blue}{(I am a bit confused about how did we calculate these numbers. The number of multiplication )} \textcolor{red}{Xinheng: the sum of multiplication and addition involved in the convolution}}
\label{fig:maxMul}
\vspace{-12pt}
\end{figure}

Figure~\ref{fig:maxMul} shows two examples of multipliers with different bitwidth configurations. For a given high bitwidth processing unit, the maximum supported throughput (multiplication and addition) of a given processing unit varies with $N$ and $K$ which are determined by the values of $p$ and $q$. For instance, when the input bitwidths of the two inputs of a multiplier are $27$ and $18$ bits (Figure~\ref{fig:maxMul27}), according to Equation~\ref{equ:slice1}, \ref{equ:slice2}, and~\ref{equ:slice3} and the required guard bits, we could obtain $S=4, N=9, K=4$ when the $p$ and $q$ are both 1-bit binary values. The maximum supported throughput of this specific multiplier is equivalent to 60 ops per cycle, which are 36 multiplications and 24 additions that are required operations for computing the convolution if all the computation is carried out in a conventional way following the 1-D convolution algorithm without HiKonv. Here, with HiKonv, all we need is one multiplication of high bitwidth multiplier with our specific slicing/packing solution. In addition, when the $p$ and $q$ are both 4 bits, the multiplier provides 8 equivalent ops per cycle (6 multiplication and 2 addition). In Figure~\ref{fig:maxMul}, we show the configurations for $p$ and $q$ from 1-bit to 8-bit, which are the common bitwidths of low-precision quantization. The principle generally applies to all bitwidths. When the inputs for the multiplier are both 32 bits, these values are further increased to 128 ops per cycle and 13 ops per cycle for 1-bit and 4-bit $p$ and $q$, as shown in the Figure~\ref{fig:maxMul32}.

% \paragraph{Batched packing}

% 1-D convolution generally has a long input sequence and a relatively shorter convolution kernel.  So we process it by maintaining the kernel data in the multiplier ($B$) and shift the data in the multiplicand ($A$) to generate new output. We then pack the new output horizontally with the previous output and segment the vertically stacked partial products. As shown in the Figure~\ref{fig:horistack}. In this way, for each of the output segments, there are only $K$ intermediate data accumulated. So the guard bit of $G_b = \lceil{log_2K}\rceil$ is set to prevent the partial results from overflow.
% \input{sections/03-formulation}
% \input{sections/04-accelerator}
\section{Evaluations}\label{sec:eva}

HiKonv is a general technique that can be adopted for both the general purpose processor and reconfigurable hardware platform. We demonstrate its efficacy on both platforms.
%We demonstrate the efficacy of our HiKonv solution on both general purpose processor and reconfigurable hardware platform.

\subsection{General Purpose Processors}
% To illustrate the efficacy of our HiKonv solution, 
We first evaluate HiKonv on both CPU-based desktop and mobile platforms with an Intel Core i7-10700K CPU and i7-10710U CPU, respectively.
% The x86-64 architecture supports both 32-bit and 64-bit multiplications. Modern CPUs are usually 64-bit processors, including the two CPUs on which we evaluate our solution. With such architecture, a 32-bit multiplication is performed with 32-bit registers as operands while 64-bit is constructed with using two 32-bit registers. 
% The 64-bit product is stored into two 32-bit registers: the upper half in one and the lower half in the other. A 64-bit multiplication is done in the same way except that the registers are 64-bit registers. Since 64-bit multiplications may produce 128-bit numbers, which are challenging to represent, we use 32-bit multiplications to ensure correctness. 
%Modern CPUs are equiped with 32-bit multiplier. Thus, without loss of generality, we use 32-bit multiplication for unsigned values, each quantized to 4-bit, in this evaluation. 
% Since we have two 32-bit operands, we have $Bit_A = Bit_B = 32$ bits according to the formulation in the Section~\ref{sec:formulation}. The bitwidths for the to-be-packed values are four bits. We have $p = q = 4$ bits. Solving the inequalities in Lemma \ref{lemma:seg}, we obtain $N = 3$, $K = 3$, and $G_b = 10$ bits.
% According to Lemma~\ref{lemma:seg}, we obtain that for 4-bit quantized convolution, $N = 3, K = 3$ and $G_b = 2, S = 10$ bits.
We measure the performance of both 1-D convolution and a DNN convolution layer. 

\begin{figure}
\begin{subfigure}{.2\textwidth}
    \begin{subfigure}{\textwidth}
        \centering
        \includegraphics[width=\linewidth]{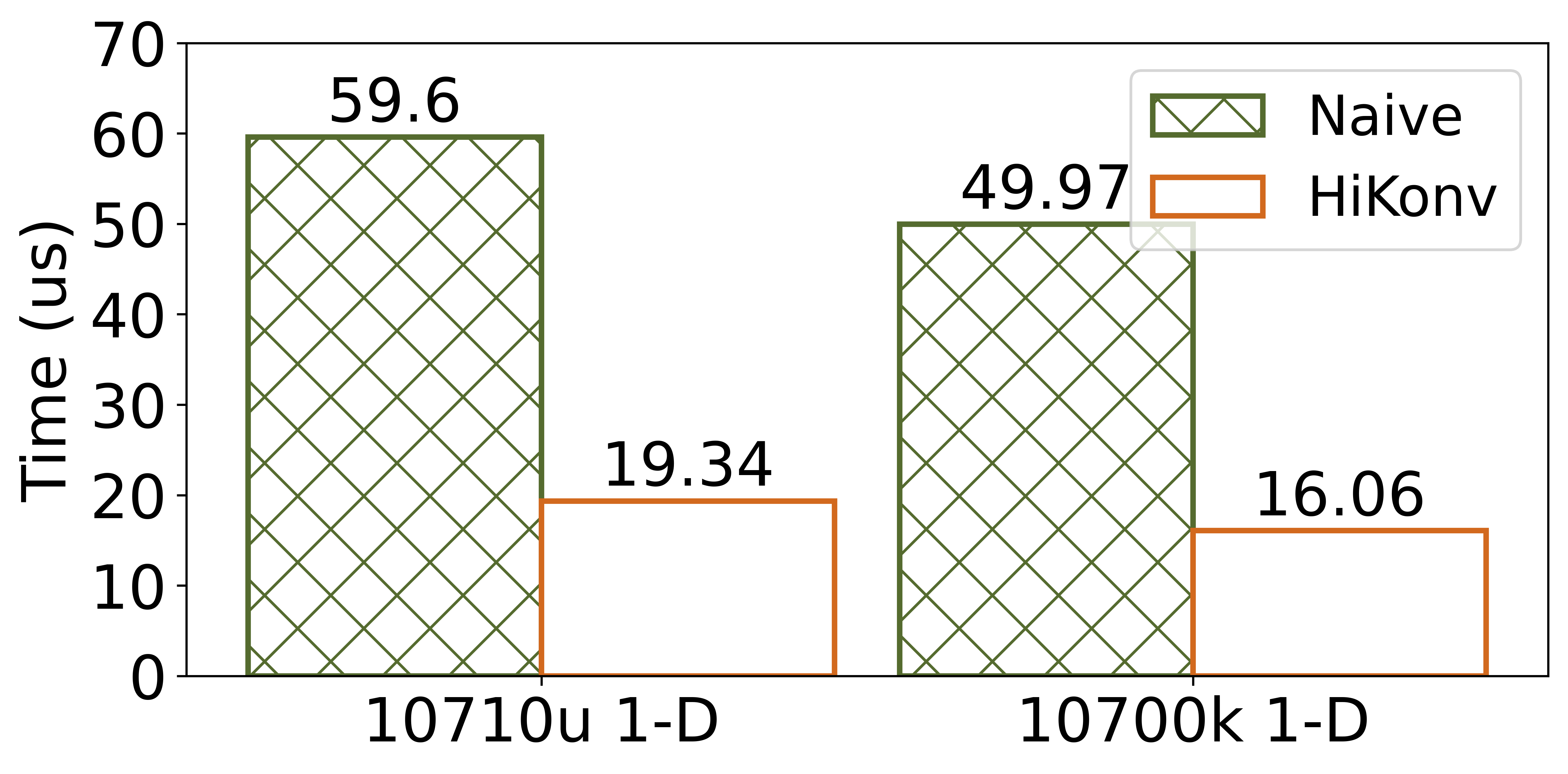}
        \vspace{-6mm}
        \caption{1-D Convolution}
        \label{fig:cpu_eval_1d}
    \end{subfigure}
    \hfill
    \vspace{1.2mm}
    \begin{subfigure}{\textwidth}
        \centering
        \includegraphics[width=\linewidth]{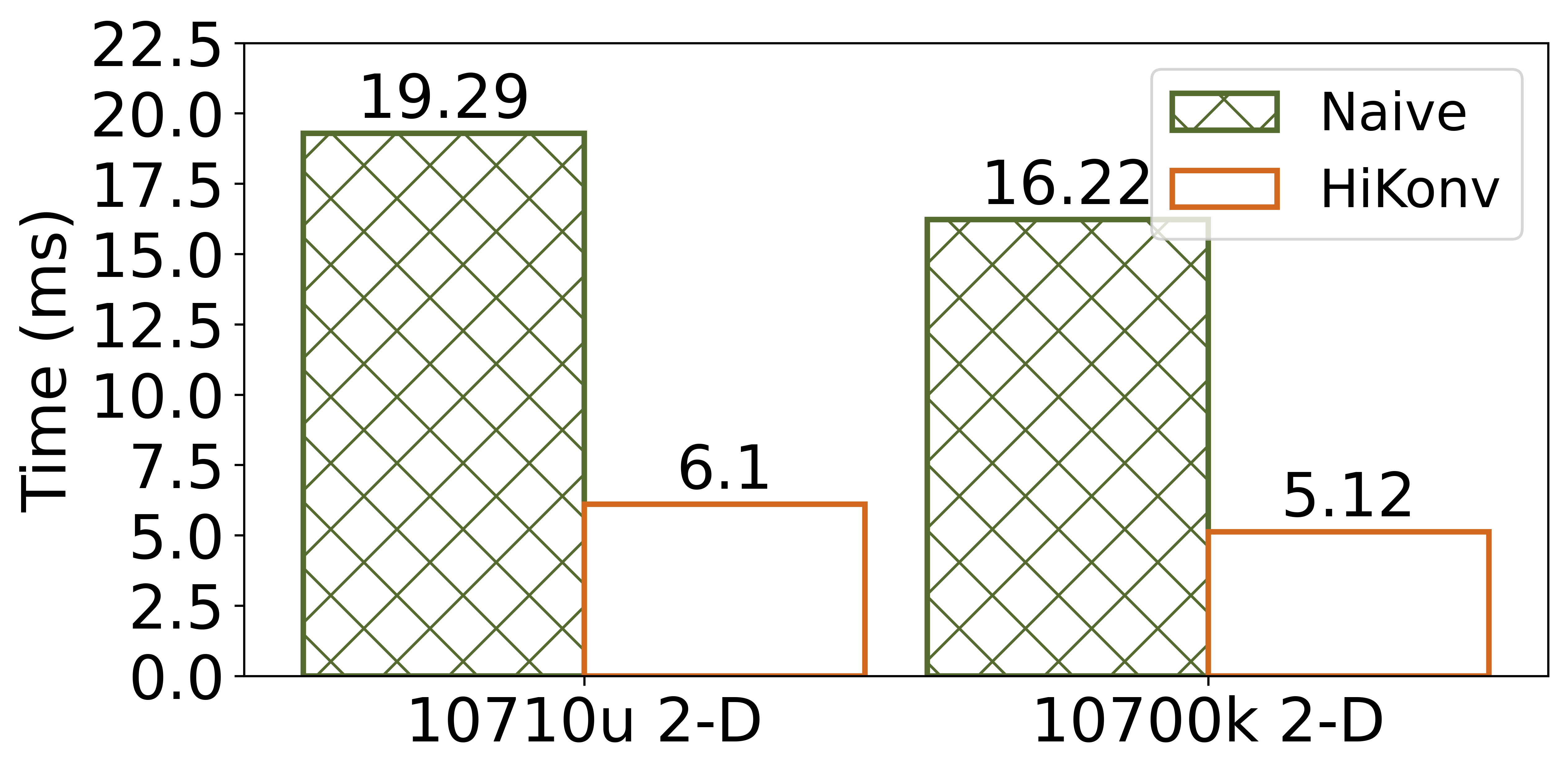}
        \vspace{-6mm}
        \caption{2-D Convolution}
        \label{fig:cpu_eval_2d}
    \end{subfigure}
\end{subfigure}
\begin{subfigure}{.27\textwidth}
    \centering
    \includegraphics[width=\linewidth]{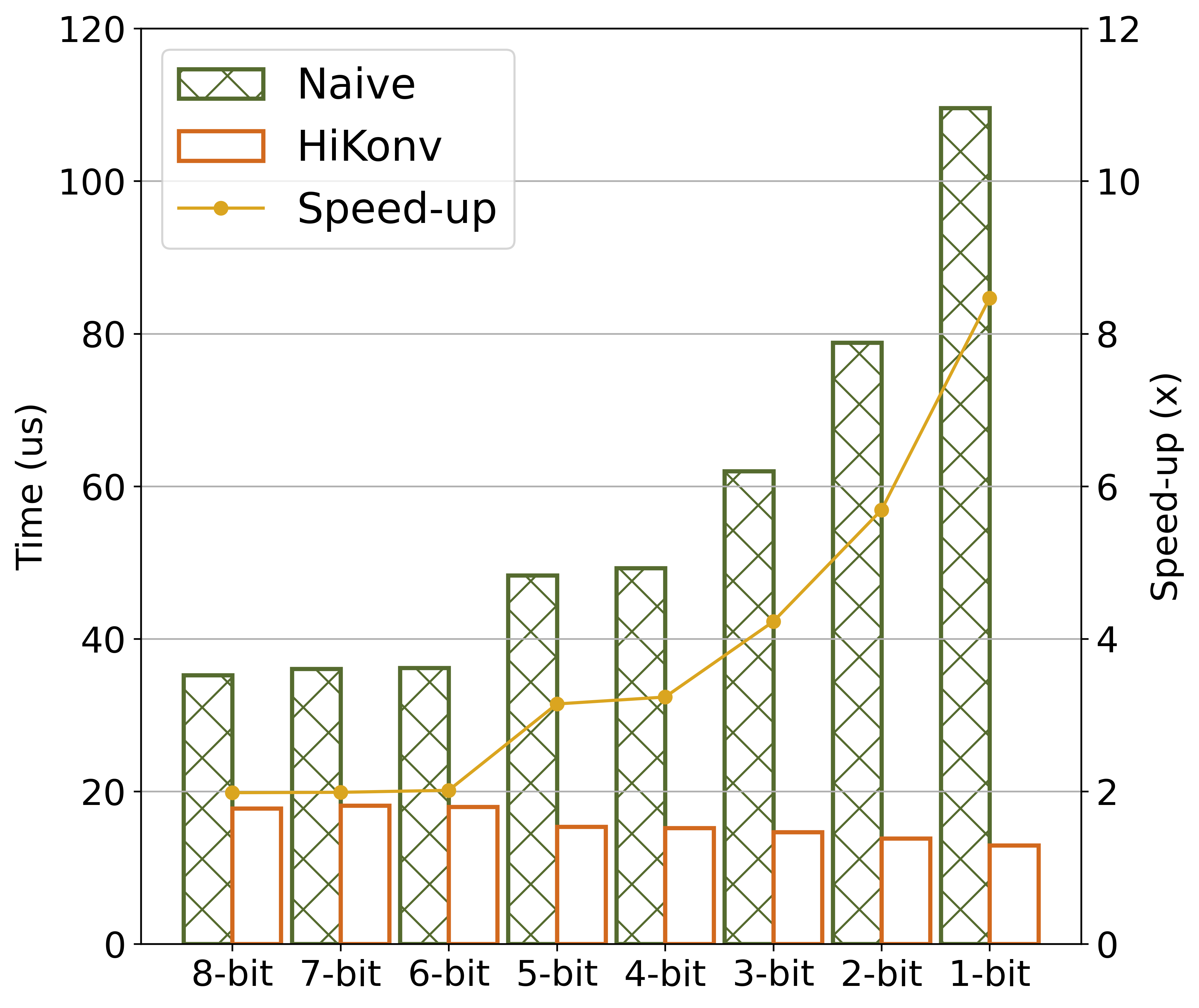}
    \vspace{-6mm}
    \caption{Speedup for different bitwidths.}
    \label{fig:all_bits}
\end{subfigure}
\caption{HiKonv Evaluation on CPU}
\vspace{-6mm}
\label{fig:cpu_eval}
\end{figure}

For 1-D convolution, the baseline implementation has 2-level nested loops. The outer loop scans through the input vector, whereas the inner loop scans through the kernel vector. %In comparison, 
% since the kernel size aligns nicely with the number of entries we pack in a 32-bit operand, 
% For the implementation of 1-D convolution with HiKonv, we only need one loop to compute the convolution. 
We adopt the horizontal stacking strategy proposed by HiKonv 1-D convolution. 
% First, we take $N$ entries from the input vector, each of which is $p$-bit in width. 
The features are packed during runtime, and kernels are packed offline before the processing starts.
% According to the segmentation that we have derived earlier, one value takes $S = 10$ bits in the 32-bit operand. We pack them into the 32-bit operand, following the guidelines provided in the previous sections, and the kernel has been packed offline for optimized performance. We then multiply together the kernel and the packed feature to produce a partial result. 
% Since we take the horizontal stacking approach, 
% we keep the partial result from the last loop and stack with the current partial result. Under this specific setting, 
For the output,
we first shift the previous partial result to the right by $S\times 2$ bits and the current partial result to the left by $S\times (N-2)$ bits. Then, we add them together to form the whole result for this loop. In the end, we take the last $S\times N$ bits 
% of the result, divide them into $S$ bits apiece, and write 
as the $N$ outputs with the corresponding indices. 
For a quantified analysis with a DNN layer, we pick the final layer of UltraNet~\cite{UltraNet2020}, which is the champion model for the DAC-SDC contest 2020 and randomly generate feature and kernel vectors. We implement the DNN layer by embedding the 1-D convolution algorithm into the 6-level nested loops that scan through the input channel, output channel, output height, output width, kernel height, and kernel width according to Theorem~\ref{thrm:convolution}. Since CPU hardware lacks bit-wise management capability, dealing with signed values can cause unnecessary overhead from intricate bit operations. While we can deploy the HiKonv solution with signed values, the hardware constraint makes such optimization less efficient than unsigned values. Since modern CPUs are equipped with 32-bit multipliers, without loss of generality, we use $A=B=32$ bits as the multiplication bitwidth, and pack $p=q=4$-bit unsigned values in each of the operands. According to Theorem~\ref{lemma:h_stack}, we obtain $N = 3, K = 3, G_b = 2$, and $S = 10$ bits. Figures \ref{fig:cpu_eval_1d} and \ref{fig:cpu_eval_2d} show the 1-D and 2-D convolution latency results, respectively. Both are compared to the baseline implementation with nested loops without our HiKonv solution.

% \begin{figure}
% \begin{subfigure}{.225\textwidth}
%     \centering
%     \includegraphics[width=\linewidth]{figures/cpu_eval_1d.png}
%     \caption{1-D Convolution.}
%     \label{fig:cpu_eval_1d}
% \end{subfigure}%
% \begin{subfigure}{.225\textwidth}
%     \centering
%     \includegraphics[width=\linewidth]{figures/cpu_eval_2d.png}
%     \caption{DNN Convolution layer.}
%     \label{fig:cpu_eval_2d}
% \end{subfigure}
% \caption{Runtime Comparison}
% \vspace{-20pt}
% \label{fig:cpu_eval}
% \end{figure}

It is clear that our HiKonv solution is about three times faster than the baseline implementations under all four combinations. 
The experimental results are slightly slower than the theoretical speed-up shown in Section~\ref{sec:formula} because of the processing overhead. Despite the reduction in loop counts and thus the total number of multiplications to generate all outputs, HiKonv has additional bit-shifting and gating operations to prepare the operands and segment the output. Since the ALU handles both multiplications and bit-wise operations, the latency of bit-wise operations are not much faster than multiplications and lead to extra overhead.

% \begin{figure}
%     \centering
%     \includegraphics[width=0.45\textwidth]{figures/all_bits.png}
%     \caption{Speedup for different bitwidths.}
%     \vspace{-20pt}
%     \label{fig:all_bits}
% \end{figure}

% % Test Image
% % Test Image Ends

We further test the performance with low-precision bitwidths from 1 to 8 bits. Assuming $p = q$, we calculate the corresponding $N$, $K$, and $G_b$ and pack the quantized values into 32-bit accordingly. Figure \ref{fig:all_bits} shows the result of the 1-D convolution with the same setting as before.
It is clear that when the bitwidth of the processed data reduces, the performance of our HiKonv solution increases because of the increased slice number. When the bitwidth is 1 bit, HiKonv solution provides a $8.6\times$ performance improvement.
% {the number of slices in the 32-bit kernel operand translates to a larger kernel size in the convolution. For example, packing three values in a 32-bit operand means the kernel size is three in the convolution. Due to the increase in the number of packs in the kernel size, the time for naive implementation increases; however, this increase does not affect the performance of our solution. With the reduction in the loop count, even though the overhead from extra bit-wise shifting and gating rises with packing, the overall performance still increases.}
% \textcolor{blue}{Our further tests show that one bit-wise operation takes roughly 1.4 nanoseconds on 10710U and 1.2 nanoseconds on 10700K, whereas one multiplication takes roughly 1.7 nanoseconds on 10710U and 1.4 nanoseconds on 10700K. As a result, even though the number of multiplications is reduced by a factor of nine. It is replaced by twelve bit-wise operations, including shifting and gating. This result gives us a strong motivation to evaluate our approach on customized hardware where bit-wise operations are considerably cheaper.
% (Junhao, can we treat the bit-wise op, shift and accumulate op as the same as multiplication op since they all use the ALU, and build the performance model for it? I feel an estimation model might be helpful to convince readers.)}

\subsection{Reconfigurable Hardware}
We also conduct the evaluation of our HiKonv solution on the Xilinx Ultra96 MPSoC platform, which is equipped with 360 DSP48E2 units and a quad-core ARM processor.
Each of the DSP48E2 has one 27-bit, one 18-bit and one 45-bit input port. It can perform one $(M_0\times M_1+Acc)$ MAC operation in one clock cycle, where $M_0, M_1, Acc$ are the inputs at the 27-bit, 18-bit, and 45-bit port, respectively.
Different from software implementation of HiKonv for general-purpose processors, with reconfigurable hardware, the input packing is conducted with small adders for each of the slices, and output segmentation is conducted by bit-wise operations. These advantages on the hardware can fully benefit the performance of our HiKonv solution.

\paragraph{Binary convolution layer}
We first evaluate the extreme case of quantized convolution 
% convolution implementation considered as the most efficient on reconfigurable hardware, 
which is the Binary Neural Networks (BNN).
A convolutional layer in a BNN takes the binary inputs for both feature maps and kernel weights, processes the convolution between them, and generates the outputs. Note that the outputs may not be binary due to the channel-wise accumulation.
We first implement a binary convolution layer with 4-bit outputs without using the DSP resources, denoted as BNN-LUT; we then replace the binary computations with our HiKonv solution with DSP, denoted as BNN-HiKonv. In comparison, we evaluate the resource utilization of these two designs under the setting of the same concurrency and same clock frequency, as shown in Table~\ref{tab:bnnEval}.
% \textcolor{blue}{A figure with the following curves, x-axis: concurrency; y1-axis: LUT usage; y2-axis: DSP usage and number of overall FPGA slices;}

\begin{table}[h]
\small
    \centering
    \caption{Comparison of Resource util. of binary convolution}
    \resizebox{\columnwidth}{!}{\begin{tabular}{c|c|c|c|c|c|c}
    \hline\hline
    \multicolumn{2}{c|}{\# of Concurrent MACs}  & 336 & 576 & 960 & 1536 & 3072\\\hline
     \multirow{1}{*}{BNN-LUT}& LUT  & 3371 & 4987 & 7764 & 12078 & 23607 \\\hline
            % & Max Freq. & 500Mhz& 500Mhz& 500Mhz& 500Mhz&500Mhz\\\hline
    \multirow{3}{*}{BNN-HiKonv} & LUT  & 2672 & 2536 & 3369 & 3587 &  9319\\ \cline{2-7}
                                & DSP   & 16   & 32   & 64   & 128  &256 \\ \cline{2-7}
                                & DSP Thro.&21 &18 &15 &12 & 12 \\ \cline{2-7}
                                & LUT/DSP & 43.7 & 76.6&68.7 & 65.4& 55.8\\ \cline{2-7}
                                % & Max Freq. & 500Mhz & 500Mhz&500Mhz & 500Mhz& 500Mhz\\ \cline{2-7}
    \hline\hline
    \end{tabular}}
    \label{tab:bnnEval}
\end{table}

Clearly, compared to BNN-LUT, the LUT usage of BNN-HiKonv is reduced. 
However, the throughput for each DSP reduces when the concurrency increases due to the reason that there is more vertical stacking, and it takes more guard bits when the concurrency increases.
The equivalent number of LUTs replaced by one DSP (LUT/DSP) varies from 43.7 to 76.6 due to the accumulation logic in the convolution operation. 
HiKonv creates opportunities to leverage DSPs for high-throughput BNN (or other low-bitwidth models) convolution computations that would help map a larger BNN with high concurrency into the same FPGA. It can also potentially increase the design's clock frequency since DSPs can run at a higher frequency than LUTs.

\paragraph{Complete model}
We apply our HiKonv solution to the entire UltraNet model~\cite{UltraNet2020} on the Xilinx Ultra96 MPSoC FPGA. 
The weight and activation of this model are quantized to 4-bit. We execute all the convolution layers on the programmable logic and the other layers on the ARM processor in the FPGA platform.
We follow the same layer architecture and system architecture as the original UltraNet design and only change the computation for convolution with our HiKonv solution.
%Only one of the 27-bit inputs and the 18-bit input are used together as the inputs to the DSP to process the convolution.
% We use additional LUTs to construct the shifting and adders following the DSP.
%Unlike the implementation of the HiKonv solution on CPUs, 
Besides using DSPs, we also use small adders and shifters constructed by LUTs, taking advantage of the flexible configuration features of the FPGA.

In addition to resources utilization, we also measure the throughput in frame-per-second (fps) and calculate the DSP efficiency in terms of Giga-operations-per-second-per-DSP (Gops/DSP) for comparison as shown in Table~\ref{tab:UltraNetimpl}.
All the testing data is first loaded into the DDR to leverage the full capacity of the accelerators in our evaluations.

% \begin{figure}
%     \centering
%     \includegraphics[width=0.5\textwidth]{figures/fpga_impl-crop.pdf}
%     \caption{Implementation on FPGA DSP.}
%     \label{fig:fpga_impl}
% \end{figure}

\begin{table}[h]
\small
\centering
% \vspace{-4pt}
\caption{UltraNet resource and performance.}
\resizebox{\columnwidth}{!}{\begin{tabular}{l|c|c|c|c}
\hline\hline
             & LUT & DSP & fps & DSP Eff. (Gops/DSP) \\ \hline
UltraNet     & 4.3k& 360 &  248 & 0.289 \\ \hline
UltraNet-HiKonv & 4.8k &  327&  401/588 & 0.514/0.753 \\ \hline\hline
\end{tabular}}
\label{tab:UltraNetimpl}
% \vspace{-4pt}
\end{table}

UltraNet-HiKonv uses more LUT resources than the original implementation due to the shifting and adding logic; however, it reduces the DSP utilization thanks to the dramatic improvements of the efficiency and the throughput of the DSPs.
The original implementation of the UltraNet uses one DSP for two 4-bit MACs that is natively supported by the synthesis tool. It only achieves 248 fps with a 0.289 Gops/DSP efficiency. 
With our HiKonv solution, the on-board implementation of UltraNet achieves 401 fps with a 0.514 Gops/DSP DSP efficiency. This significant improvement is achieved under the constraint that the software execution on the ARM core is not fast enough to feed the input data to the FPGA accelerator to process, even with our best software optimization of multi-threading and data buffering. If this ARM core bottleneck is removed, the UltraNet-HiKonv accelerator can reach an even higher performance of 588 fps with the DSP efficiency of 0.753 Gops/DSP.
\section{Related Works}\label{sec:backnrelate}

% \textcolor{blue}{In general, I feel, there is no existing solution same as ours, so we could present in this way, to emphasize the lack of the support.}

% \paragraph{1. Software solutions only deal with ....}
% \textcolor{purple}{Prakhar: Add in the survey about the software solutions that you have collected here.}

Existing solutions for low-bitwidth arithmetic \cite{reuther2020survey} build their own computation units based on the inputs \cite{sharma2018bit,ryu2019bitblade,shin2018dnpu,lee2018unpu,sharify2018loom,pirdsp} and benefit from the control flexibility down to a single bit.
% Such an extreme level of fragmentation allows them to customize for variable bitwidth and focus on only relevant computations.
Prior work for accelerating DNN inference has also incorporated low-bitwidth computations. Tensor processing units (TPUs) introduce 16-bit \textit{bfloats} \cite{wang2019bfloat16}, and mobile GPUs and other edge devices now support 8-bit computations. 
However, these improvements focus only on homogeneous arithmetic requirements and do not allow flexible arithmetic computations with varied bitwidths. Therefore, when processing data with a bitwidth different from its targeted bitwidth, it either leaves some precision unused with wasted resources or hurts the efficiency of the overall process.

% Bottom-up methods generally require long design cycle due to the change of the hardware and lose the flexibility to tackle the change of the data types.

% Top-down methods, on the other hand, attempt to pack short bitwidth values into longer words provided by the hardware and use existing computing units for low-level arithmetic \cite{AVXforquant, stojanov2018fast,ottavi2020mixed,garofalo2020pulp,lai2018cmsis}. 
There are methods that simply pack short bitwidth values into longer words and attempt to incorporate additional computations using the existing unit through bit shifting and packing~\cite{AVXforquant, xilinxint4, ottavi2020mixed,garofalo2020pulp,lai2018cmsis} to further improve processing efficiency.
However, those studies are ad-hoc and do not fully leverage the hardware's capability as HiKonv does. Moreover, there are no theoretical studies to guide the flexible management of low-bitwidth quantized data. Our work fills the gap of processing low-bitwidth data under theoretical guidance for the best computation efficiency and throughput on either existing hardware architecture or any bit-efficient processing units in the future.

\section{Conclusion and discussion}\label{sec:conclusion}

In this paper, we present HiKonv, a general technique with theoretical guarantees for using a single multiplier unit to process multiple low-bitwidth convolution operations in parallel for significantly higher computation throughput  with flexible bitwidths.
It is able to support convolutions in DNNs and achieves the highest possible throughput for quantized convolution with novel bit-wise management and computation. 
% However, our HiKonv solution still have the space for improvement which will be our future work direction. Currently, our work only support stride 1 convolution with high efficiency. When dealing with signed convolution, it still requires additional adder or addition operation  to further formalize the output as required.
% HiKonv adopts novel bit management and computation and
% fills the current missing gap for efficient processing of low-bitwidth quantized convolution on existing high precision arithmetic units. 
As a demonstration of its general applicability and benefits, we show that HiKonv has achieved $3.17\times$ throughput improvement on CPU and $2.37\times$ and $2.61\times$ throughput and DSP efficiency improvements for the DAC-SDC 2020 champion model on FPGA.
HiKonv suits for both software and hardware optimizations and provides new opportunities for future hardware designs for efficient DNN processing.

\section*{\sc Acknowledgement}
This work is supported in part by the IBM-Illinois Center for Cognitive Computing Systems Research (C3SR), the National Research Foundation, Prime Minister's Office, Singapore under its Campus for Research Excellence and Technological Enterprise (CREATE) programme, and Xilinx Adaptive Compute Cluster at University of Illinois at Urbana-Champaign.

\setstretch{0.75}
\newcommand{\BIBdecl}{\setlength{\itemsep}{0.25em}}
\bibliographystyle{IEEEtran}
% \bibliography{IEEEabrv, ref}
\bibliography{IEEEabrv, ref_full}

%\begin{thebibliography}{9}
%	\footnotesize
%	\bibitem{key}
%	I. M. Author,
%	``Some related article I wrote,''
%	{\em Some Fine Journal}, vol. 17, pp. 1--100, 1987.
	%
%	\bibitem{baz}
%	A. N. Expert,
% 	{\em A Book He Wrote,}
% 	His Publisher, 1989.

% 	\bibitem{unpub}
% 	M. Smith,
% 	``Title of paper optional here,''
% 	unpublished.

% 	\bibitem{inpress}
% 	K. Rose,
% 	``Title of paper with only first word capitalized,''	% bug fixed by M. Imai
% 	in press.

% 	\bibitem{trans}
% 	T. Murayama,
% 	``Title of paper published in translation journals,''	% bug fixed by M. Imai
% 	{\em Some English Journal}, vol. 17, pp. 1--100, 1995.	% bug fixed by M. Imai
% 	({\em Original Foreign Journal, vol. 1, pp. 100-200, 1993}.)	% ditto

% \end{thebibliography}

\end{document}